\documentclass[aps,floats,prl,twocolumn]{revtex4}
\usepackage{amsfonts}
\usepackage{amsmath}
\usepackage{amssymb}
\usepackage[final,dvips]{graphicx}

\begin{document}

\title{Mixed state of a dirty two-band superconductor: application to MgB$_{2}$}
\author{A.\ E.\ Koshelev}
\affiliation{Materials Science Division, Argonne National
Laboratory, Argonne, Illinois 60439}
\author{A.\ A.\ Golubov}
\affiliation{Department of Applied Physics, University of Twente,
7500 AE Enschede, The Netherlands}

\date{\today}

\begin{abstract}
We investigate the vortex state in a two-band superconductor with
strong intraband and weak interband electronic scattering rates.
Coupled Usadel equations are solved numerically and the
distributions of the pair potentials and local densities of states
are calculated for two bands at different values of magnetic
fields. The existence of two distinct length scales corresponding
to different bands is demonstrated. The results provide
qualitative interpretation of recent STM experiments on vortex
structure imaging in MgB$_{2}$.
\end{abstract}

\maketitle

\vspace{-0.2in}%
A very peculiar feature of the recently discovered
superconductor MgB$_2$ \cite{Akimitsu}  is the multigap nature of
the superconducting state.
The possibility of such a state was first predicted in
\cite{Suhl,Moskal} for a multiband superconductor with large
disparity of the electron-phonon interaction for the different
Fermi-surface sheets. Various aspects of multiband
superconductivity, in particular the role of impurity scattering,
were discussed theoretically in
\cite{SungJPCS67,SchopohlSSC77,Kre,GolubMazPRB97}. For MgB$_2$,
the two-band model was first suggested in \cite{LiuPRL01,Shulga}.
On the basis of first-principles calculations of the electronic
structure and the electron-phonon interaction, it was argued that
superconductivity in this compound resides in two groups of bands:
the group of two strongly superconducting
$\sigma$-bands and the group of two weakly superconducting
$\pi$-bands. Quantitative predictions for $T_c$, energy gaps,
specific heat \cite{Choi,Golub} and tunneling \cite{Brink} were
made recently for MgB$_2$.

Signature of two energy gaps was observed in Nb doped SrTiO$_3$
\cite{Bednorz}.
%
But to date, only in MgB$_2$
existence of two distinct gaps has been most clearly demonstrated.
A large number of experimental data, in particular
tunneling \cite{GiubileoPRL01,IavaronePRL02} and point contact
measurements \cite{RubioPRL01,SzaboPRL01,SchmidtPRL01,Gonnelli}
and heat capacity measurements \cite{BouquetPRL01}, directly
support the concept of a double gap MgB$_2$.
It was argued in Ref.\ \cite{Mazin02} that
the unexpectedly weak correlation between $T_c$ and the
resistivity can be reconciled with the two-band model, if one
assumes that the interband impurity scattering remains weak even
in samples with strong intraband impurity scattering in the
$\pi$-band.

Two-band superconductivity in MgB$_{2}$ offers new interesting
physics.
For example, it was demonstrated that the anisotropies of the
upper critical fields and the London penetration depths are
different and have opposite temperature dependencies \cite{Kogan}.
Recently, the c-axis Abrikosov vortex structure in MgB$_{2}$ was
studied by STM \cite{Eskildsen02}. Several important observations
have been made: large vortex core size compared to estimates based
on $H_{c2}$, the absence of zero-bias singularity in the core and
the rapid suppression of
the apparent tunneling gap
by magnetic fields much smaller than $H_{c2}$.
Important property, that is essential for understanding these
findings, is that c-axis tunneling in MgB$_{2}$ probes mainly the
weakly superconducting $\pi$-band \cite{IavaronePRL02}.

In this paper, we provide a quantitative model for the vortex
structure in a two-band superconductor. We demonstrate the
existence of two different spatial and magnetic field scales,
consistent with the data in \cite{Eskildsen02}.

We consider a two-band superconductor with weak interband impurity
scattering and rather strong intraband scattering rates exceeding
the corresponding energy gaps (dirty limit). In this case the
quasiclassical Usadel equations \cite{Usadel} are applicable
within each band. The vortex structure in single-band dirty
superconductors was studied extensively in the framework of the
Usadel equations \cite{KramerJLTP74,GolubKuprJLTP88}. To describe
the mixed state in the considered case, one can generalize the
approach \cite{KramerJLTP74,GolubKuprJLTP88} and write down the
system of coupled Usadel equations
\begin{eqnarray}
&\omega F_{\alpha}\! -\!\frac{\cal{D}_{\alpha}}{2}\left[
G_{\alpha}(\mathbf{\nabla}\!-\! \frac{2\pi
i}{\Phi_{0}}\mathbf{A})^{2}F_{\alpha}\!-\!F_{\alpha}\mathbf{\nabla
} ^{2}G_{\alpha}\right]\! =\!\Delta_{\alpha}G_{\alpha}
\label{UsadelFG}\\
&\Delta_{\alpha}= 2\pi
T\sum_{\beta,n}\Lambda_{\alpha\beta}F_{\beta } \label{SelfCons}
\end{eqnarray}
where $\alpha=1,2$ is the band index, $\hat {\Lambda}$ is the
matrix of effective coupling constants (to be defined below),
$\mathcal{D}_{\alpha}$ are diffusion constants, which determine
the coherence lengths $\xi_{\alpha}=\sqrt { \mathcal{D}_{\alpha}/2
\pi T_{c}}$, $G_{\alpha}$ and $F_{\alpha}$ are normal and
anomalous Green's functions
connected by normalization condition
$G_{\alpha}^2+F^*_{\alpha}F_{\alpha}=1$,
$\Delta_{\alpha}$ is the pair potential and $\omega=(2n+1)\pi T$
are Matsubara frequencies. Bearing in mind the application to
MgB$_{2}$, in our notations index 1 corresponds to $\sigma$-bands
and index 2 to $\pi$-bands.

Note that in the considered case of weak interband scattering the
Green's functions in different bands are coupled only indirectly,
via the self-consistency equation (\ref{SelfCons}). As will be
shown below, this fact leads to the existence of two different
length scales in different bands and, as a consequence, two
magnetic field scales appear
which are directly accessible experimentally.
Such a situation has never existed in the field of vortex physics.
This is in contrast to the usual proximity effect in real space
(e.g. $N/S$ multilayers), where different energy and length scales
exist in spatially separated $N$, $S$ layers.

Let us study the case when magnetic field is oriented
along c-axis
Further, we neglect in-plane anisotropy and adopt a circular cell
approximation
for the vortex unit cell \cite{KramerJLTP74} (see inset in Fig.\
\ref{Fig-N(E,r)} ). We also assume a large Ginzburg-Landau
parameter $\kappa$, $\kappa\gg1$ (this assumption is fulfilled in
MgB$_{2}$)
and consider magnetic fields significantly larger than the lower
critical field, so that we can neglect variations of the magnetic
field.
To facilitate the analysis, we introduce reduced variables: we
will use $\pi T_{c}$ as a unit of energy, and
$\xi_{1}=\sqrt{\mathcal{D}_{1}/2 \pi T_{c}}$ as a unit of length.
The distribution of superfluid momentum within the unit cell of
the lattice is then given by
\begin{equation}
p=1/r-r/r_{c}^{2},\ r_{c}^{2}=H_{1}/H,\ H_{\alpha}\equiv T_{c}\Phi
_{0}/ \mathcal{D}_{\alpha}   \label{Supermom}
\end{equation}
where $r$ is the distance from the center of a vortex core.

Using $\theta-$parametrization ($F_{\alpha}=\sin\theta$,
$G_{\alpha}=\cos\theta_{\alpha}$) the Usadel equations and the
self-consistency conditions can be rewritten in the form
\begin{eqnarray}
\partial_{r}^{2}\theta_{\alpha}&+&\frac{1}{r}\partial_{r}\theta_{\alpha}-p^{2}
\cos\theta_{\alpha}\sin\theta_{\alpha}\nonumber\\
&+&k_{\alpha}^{2}\left(
\Delta_{\alpha}\cos\theta_{\alpha}-\omega\sin\theta_{\alpha}\right)
=0 \label{Usadel-Reduced}
\end{eqnarray}
\vspace{-0.2in}
\begin{subequations}
\begin{align}
&W_{1}\Delta_{1}\!-W_{12}\Delta_{2}\!
=\!2t\sum_{\omega>0}\!\left( \sin\theta _{1}-
\frac{\Delta_{1}}{\omega}\right)\! +\Delta_{1}\ln\frac{1}{t}
\label{SelfCons-Reduced} \\
&\!-\!W_{21}\Delta_{1}\!+\!W_{2}\Delta_{2}\!
=\!2t\sum_{\omega>0}\!\left(\! \sin\theta _{2}-
\frac{\Delta_{2}}{\omega}\!\right)\! +\!\Delta_{2}\ln\frac{1}{t}
\end{align}
\end{subequations}
with $k_{1}^{2}=1$, $k_{2}^{2}=\mathcal{D}_{1}/\mathcal{D}_{2}$, $
\omega=t(2n+1)$, $t=T/T_{c}$. The matrix $W_{\alpha\beta}$ is
related to the coupling constants $\Lambda_{\alpha\beta}$ as
\begin{align}
W_{1} & \!=\frac{-A+\sqrt{A^{2}+\Lambda_{12}\Lambda_{21}}}{\mathrm
{Det}},\ W_{2}\!=\frac{
A+\sqrt{A^{2}+\Lambda_{12}\Lambda_{21}}}{\mathrm{Det}},   \notag\\
W_{12} & =\Lambda_{12}/\mathrm{Det},\
W_{21}=\Lambda_{21}/\mathrm{Det}  \label{MatrixLambda}
\end{align}
where $A=(\Lambda_{11}-\Lambda_{22})/2,$
$\mathrm{Det}=\Lambda_{11}\Lambda_{22}- \Lambda_{12}\Lambda_{21}$
\cite{Note-SingleOP}.
Note that only three constants are
independent since $W_{1}W_{2}=W_{12}W_{21}.$

Partial local densities of states (DoS)
$N_{\alpha}(\varepsilon,r)$, which are accessible in tunneling
experiments, can be obtained from $\Theta_{\alpha }(\omega,r)$
using analytic continuation
\begin{equation}
N_{\alpha}(\varepsilon,r)=\operatorname{Re}\left[
\cos\Theta_{\alpha} (i \omega\rightarrow
\varepsilon+i\delta,r)\right] \label{DOS}
\end{equation}

The above set of equations (\ref{Supermom})-(\ref{DOS}) fully
defines vortex core structure in a diffusive two-band
superconductor. In general, numerical solution is required to
determine the behavior of the pair potentials and DoS in both
bands. The problem simplifies near the upper critical field when
Eqs.\ (\ref{Usadel-Reduced}) can be linearized
\begin{equation}
\partial _{r}^{2}\theta _{\alpha }+\frac{1}{r}\partial _{r}\theta _{\alpha
}-\left( 1/r-r/r_{c}^{2}\right) ^{2}\theta _{\alpha }-k_{\alpha }^{2}\omega
\theta _{\alpha }=-k_{\alpha }^{2}\Delta _{\alpha }.  \label{Usadel-Hc2}
\end{equation}
These equations have exact solution \cite{KramerJLTP74}
\[
\Delta _{\alpha } =\Delta _{0\alpha }r\exp \left(
-r^{2}/2r_{c}^{2}\right), \theta _{\alpha } =\theta _{0\alpha
}r\exp \left( -r^{2}/2r_{c}^{2}\right) ,
\]
giving the relation
\begin{equation}
\theta _{0\alpha }=\frac{\Delta _{0\alpha }}{2/k_{\alpha
}^{2}r_{c}^{2}+\omega }.
\end{equation}
Substituting this result into the self-consistency equations we obtain
\begin{subequations}
\begin{align}
W_{1}\Delta _{01}-W_{12}\Delta _{02}& =\Delta _{01}\left( \ln
\frac{1}{t}
-g\left( \frac{H}{tH_{1}}\right) \right)   \label{SelfCons-Hc2} \\
-W_{21}\Delta _{01}+W_{2}\Delta _{02}& =\Delta _{02}\left( \ln
\frac{1}{t} -g\left( \frac{H}{tH_{2}}\right) \right)
\end{align}
\end{subequations}
where $g(x)\equiv \psi (1/2+x)-\psi (1/2)$ and $\psi (x)$ is a
digamma function. This gives the equation for $ H_{c2}$
\begin{equation}
\frac{W_{1}}{\ln \frac{1}{t}-g\left( \frac{H}{tH_{1}}\right)
}+\frac{W_{2}}{ \ln \frac{1}{t}-g\left( \frac{H}{tH_{2}}\right)
}=1  \label{Eq-Hc2}
\end{equation}
and relation between $\Delta _{01}$ and $\Delta _{02}$ near $H_{c2}$
\begin{equation}
\Delta _{02}=\frac{W_{21}\Delta _{01}}{W_{2}+\ln
\frac{1}{t}-g\left( \frac{H }{tH_{2}}\right) }
\end{equation}
In the single band case, it follows directly from
Eq.(\ref{SelfCons-Hc2}) that the upper critical field $H_{c2}^{s}$
is given by the standard Maki - de Gennes equation
\begin{equation}
\ln (1/t)=g\left [H_{c2}^{s}/(tH_{1})\right].
\label{Eq-Hc2-SingleBand}
\end{equation}
\begin{figure}
[ptb]
\begin{center}
\includegraphics[
width=3in ] {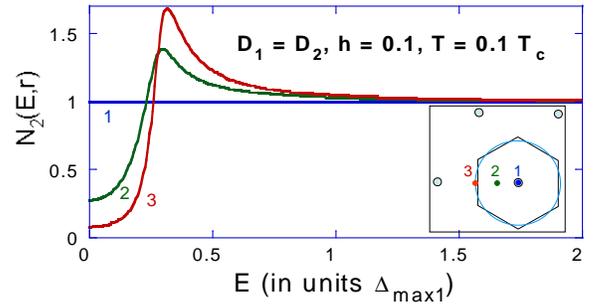} \caption{Local density of state for
$\pi$-band at different points of vortex lattice unit cell at
$h=0.1$ for $\mathcal{D}_{1}=\mathcal{D}_{2}=1$. Inset illustrates
the circular cell approximation and shows points at which the
spectra are taken.} \label{Fig-N(E,r)}
\end{center}
\vspace{-0.25in}
\end{figure}

\textbf{Application to MgB$_{2}$}. The electron-phonon interaction
in MgB$_{2}$ was calculated from first principles in
\cite{LiuPRL01,Choi,Golub}. In Ref. \cite{Golub} the matrices of
the electron-phonon coupling constants $\lambda_{ij}$ and the
renormalized Coulomb pseudopotentials $\mu _{ij}^{\ast}$ were
derived for the effective two-band model. In this paper we will
use these results and define the effective constants
$\Lambda_{ij}=\lambda_{ij}-\mu_{ij}^{\ast}$ in the weak coupling
model, neglecting the strong-coupling corrections, which is a
reasonable approximation for our purpose.
The corresponding numerical values are \cite{Golub}
\vspace{-0.05in}
\begin{equation}
\!\Lambda_{11}\!\approx0.81,\ \Lambda_{22}\!\approx0.278,\
\Lambda_{12}\!\approx0.115,\ \Lambda_{21}\!\approx0.091,
\vspace{-0.05in}
\end{equation}
from which we obtain values of $W_{\alpha\beta}$ used in numerical
calculations, \vspace{-0.05in}
\begin{equation}
\!W_{1}\!\approx0.088,\ W_{2}\!\approx2.56,\
W_{12}\!\approx0.535,\ W_{21}\!\approx0.424. \vspace{-0.05in}
\end{equation}
With fixed coupling constants the overall behavior is determined
by the ratio of the diffusion constants
$\mathcal{D}_{1}/\mathcal{D}_{2}$, which is not known at present
and may depend on the type of scatterers.
Available estimates of scattering rates \cite{YelandPRL02} suggest
that typically $\mathcal{D}_{1}\gtrsim \mathcal{D}_{2}$. However
we found that to describe the experimental vortex core structure
we have to take $\mathcal{D}_{1}< \mathcal{D}_{2}$ (as will be
discussed below, this  apparent contradiction can be resolved by
going to $\xi_1/\xi_2$ ratio). Therefore, we present calculations
for two values of the ratio $\mathcal{D}_{1}/\mathcal{D}_{2}$:
$\mathcal{D}_{1}/ \mathcal{D}_{2}=0.2$ and
$\mathcal{D}_{1}/\mathcal{D}_{2}=1$.

\begin{figure}
[ptbptb]
\begin{center}
\includegraphics[clip, width=3.4in ]
{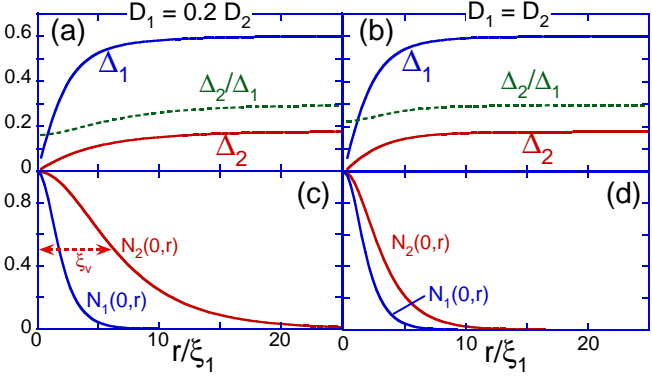} \vspace{-0.1in}\caption{Spatial dependencies of
pair potentials ((a) and (b)) and partial DoS at $E=0$ ((c) and
(d)) for isolated vortex for two ratios $D_1/D_2$: $D_1= 0.2D_2$
and $D_1= D_2$} \label{Fig-IsolVortex}
\end{center}
\vspace{-0.25in}
\end{figure}

The magnetic field is measured with respect to the single-band
upper critical field of the $\sigma $-band,
$h\equiv H/H_{c2}^{s}(t)$,
where $H_{c2}^{s}$ is given by equation
(\ref{Eq-Hc2-SingleBand}).
For the case $W_1\ll W_2$ realized in MgB$_2$, the upper critical
field is mainly determined by the strong band. Small correction
due to the weak band can be found from Eq.\ \ref{Eq-Hc2} using
expansion with respect to the small parameter $S_{12}\equiv
W_1/W_2$. In particular we found very simple expressions for the
slope of $H_{c2}$ at $T_c$ and $H_{c2}(0)$:
\begin{align*}
\frac{dH_{c2}}{dT}&\approx\frac{dH_{c2}^{s}}{dT}\left (
1+S_{12}\frac{H_{2}-H_{1}}{H_{2}}\right )\\
H_{c2}(0) &\approx H_{c2}^{s}(0)\left(  1+S_{12}\ln\left(
H_{2}/H_{1} \right)  \right).
\end{align*}

With the above parameters, we numerically solved equations
(\ref{Usadel-Reduced}) and (\ref{SelfCons-Reduced}) for different
magnetic fields. Fig. \ref {Fig-N(E,r)} shows an example of local
DoS for the $\pi $-band at different points of the vortex unit
cell. One can see that in the center of the core
there is no zero-energy peak in the DoS in the core usually
observed in clean superconductors \cite{Renner}. This property is
a consequence of the dirty limit
in the $\pi-$band.
As one can expect, the most pronounced dependence on energy is
observed at the boundary of the vortex unit cell (curve 3 in Fig.
\ref {Fig-N(E,r)}). One can see that the DoS is peaked at an
energy about 3 times smaller than $\Delta_{max1}$. This peak
corresponds to the small energy gap in the second band.
\begin{figure}
[ptbptbptb]
\begin{center}
\includegraphics[clip, width=3.4in]
{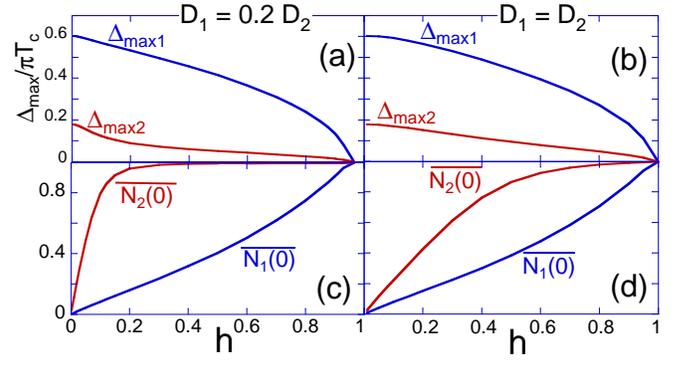} \vspace{-0.1in}\caption{Field dependencies of maximum
pair potentials ((a) and (b)) and averaged DoS at $\epsilon=0$
((c) and (d)) for two ratios $D_1/D_2$: $D_1= 0.2D_2$ and $D_1=
D_2$ } \label{Fig-FieldDep}
\end{center}
\vspace{-0.1in}
\end{figure}
\begin{figure}
[ptbptbptbptb]
\begin{center}
\includegraphics[clip, width=2.4in]
{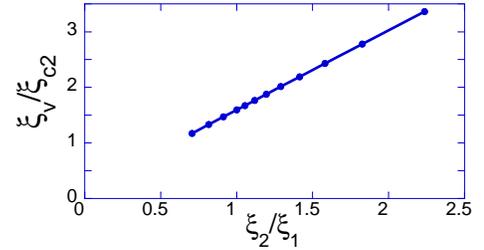}\vspace{-0.1in} \caption{ Ratio of the apparent
vortex size $\xi_v$ as defined in Fig.\
\protect\ref{Fig-IsolVortex}c to the coherence length $\xi_{c2}$
extracted from the upper critical field, $\xi_{c2}\equiv \sqrt{
\Phi_0/(2\pi H_{c2})}$ plotted vs ratio of the typical length
scales in two bands, $\xi_2/\xi_1\equiv
\sqrt{\mathcal{D}_2/\mathcal{D}_1}$.
According to Ref.\ \protect\cite{Eskildsen02} for MgB$_2$ single
crystal the ratio $\xi_v/\xi_{c2}$ is around 3.
}\label{Fig-VortexSize}
\end{center}
\vspace{-0.25in}
\end{figure}

We study structure of an isolated vortex by solving the Usadel
equations at very small field ($h\!=\!0.002$). Fig.\
\ref{Fig-IsolVortex}a,b shows the spatial dependence of the pair
potentials, $\Delta_{1}(r)$ and $\Delta_{2}(r)$, and their ratio
for $t\!=\!0.1$ for two cases: $D_{1}\!=\! 0.2D_{2}$ and
$D_{1}\!=\! D_{2}$. Fig.\ \ref{Fig-IsolVortex}c,d shows the DoS at
zero energy, $ N_{1}(0,r)$ and $N_{2}(0,r)$. One can see that in
the case of $D_{1}\!=\! 0.2D_{2}$ the pair potential and the DoS
in the $\pi$-band demonstrate qualitatively different behavior.
The pair potentials approach their bulk values $\Delta_{\alpha,0}$
at the length scale set by the strong band. $ \Delta_{1}$ reaches
half of $\Delta_{1,0}$ at $r\!=\!2.15\xi_{1}$ and $\Delta _{2}$
reaches half of $\Delta_{2,0}$ at a somewhat larger length scale,
$r\!=\!3.44\xi_{1}$. $\pi$-band DoS, $N_{2}(0,r)$, has
significantly longer range: it drops to $0.5$ at
$r\!=\!6.35\xi_{1}$.
Though the above numbers correspond to the specific choice of
parameters for the coupling matrix $\Lambda_{ij}$, the large core
size in the weakly superconducting band is the general property of
a two-band superconductor.

The two typical sizes of the isolated vortex determine the two
typical field scales. Fig.\ \ref{Fig-FieldDep} shows the field
dependence of the maximum values of the pair potentials at the
boundary of vortex unit cell ((a) and (b)) and DoS at $\epsilon=0$
averaged over the unit cell ((c) and (d)). One can see that for
the case of $D_{1}= 0.2D_{2}$ the average DoS in the $\pi$-band
reaches its normal value at fields considerably smaller than the
upper critical field.

Recently, the c-axis vortex structure in MgB$_{2}$ single crystals
was measured by STM \cite{Eskildsen02}. Most strikingly, it was
observed that the spatial extension of the vortex core was a few
times larger than the length $\sim$10 nm estimated from $H_{c2}$.
Our model provides an natural explanation of this fact. One can
see from Fig.\ \ref{Fig-VortexSize} that the apparent vortex size
can indeed exceed the size estimated from the $H_{c2}$. The
magnitude of the enhancement depends on the ratio of the diffusion
constants in the two bands. As follows from numerical
calculations, the apparent vortex size $\xi_v$ is approximately
given by the expression $\xi_v = 2.7 \xi_2 + 0.3 \xi_1$. The low
energy peak (around 2.2 meV) in the region between the vortices
and its rapid suppression by magnetic field is also explained by
our model.

The measured value of $\xi_v/\xi_{c2}=3$ corresponds in our model
to the ratio $\xi_2/\xi_1=2$. We can not make a quantitative
comparison between the measured and calculated values of $\xi_v$,
since scattering parameters in different bands for MgB$_{2}$
single crystals of Ref.\ \onlinecite{Eskildsen02} are not known.
Moreover, from the available data on resistivity and de Haas - van
Alphen effect \cite{YelandPRL02}, it follows that the
$\sigma$-band in MgB$_{2}$ single crystals is in the clean limit.
At the same time, the $\pi$-band, probed by c-axis tunneling, is
moderately dirty,
which is consistent with theoretical estimates \cite{Mazin02}.
Due to the increase of the effective coherence
length at low energies \cite{Usadel,KramerJLTP74}, the dirty limit
condition in the $\pi$-band is certainly satisfied in the energy
range $E<\Delta_{max2}$. This is consistent with the absence of
localized states in the vortex core as claimed in
Ref.\cite{Eskildsen02}.

Our results should still be qualitatively applicable to MgB$_{2}$,
even if the $\sigma$-band is in the clean limit.
Indeed, if we focus on the DoS in the $\pi$-band, which is in the
dirty limit, then the Usadel equation for Green's function for
this band is still valid. The only extra input we need is the
coordinate dependence of the pair potential $\Delta_2$ in the
$\pi$-band which is coupled to the pair potential $\Delta_1$ in
the $\sigma$-band. The shape of the coordinate dependence of
$\Delta_1$ does not depend much on the degree of dirtiness, only
the scale of this dependence changes. Therefore, our result can be
considered as phenomenological if we define $\xi_1$ as the typical
scale of change of $\Delta_1$. Moreover, $\xi_1$ has small weight
in the dependence $\xi_v (\xi_2,\xi_1)$.

In conclusion, the vortex core structure in a dirty two-band
superconductor with weak interband scattering is studied
theoretically. The distributions of the order parameters and local
DoS reveal two different spatial scales for the two bands, in
qualitative agreement with recent STM experiments on MgB$_{2}$.
This further supports the two band model in MgB$_{2}$ and also
provides an interesting new type of vortex core structure.

We acknowledge valuable discussions with A.Brinkman, O.V.Dolgov,
I.I.Mazin, M. Iavarone, and G. Karapetrov. In Argonne this work
was supported by the U.S. DOE, Office of Science, under contract
\# W-31-109-ENG-38. 


\begin{thebibliography}{99}
\bibitem{Note-SingleOP} Note that in many theoretical papers the
model of "factorizable" interactions is used (see, e.\ g., Ref.\
\protect\onlinecite{Kogan} ), $\Lambda_{\alpha \beta}=g_\alpha
g_\beta$. This corresponds to the limit $W_{\alpha \beta}
\rightarrow \infty$ with the fixed ratio
$W_{21}/W_2=W_{1}/W_{12}$. In this limit the gap ratio
$\Delta_2/\Delta_1$ is rigidly fixed, leading to the model with a
single order parameter distributed over the Fermi surface. We
should note that there is no physical justification for this model
and it has to be considered no more than a toy model, even though
in many cases it catches essential physics.
\end{thebibliography}
\end{document}